\documentclass[fleqn,12pt,twoside]{article}
\usepackage[headings]{espcrc1}
\readRCS
$Id: espcrc1.tex,v 1.2 2004/02/24 11:22:11 spepping Exp $
\usepackage{graphicx}
\usepackage[figuresright]{rotating}
\usepackage{epsfig}

\newcommand{\AmS}{{\protect\the\textfont2
  A\kern-.1667em\lower.5ex\hbox{M}\kern-.125emS}}
\title{System and rapidity dependence of baryon to meson ratios at RHIC}
\author{Eun-Joo Kim\address[MCSD]{University of Kansas, Lawrence, Kansas 66045, USA} 
	for the BRAHMS Collaboration}
\runtitle{System and rapidity dependence of baryon to meson ratios at RHIC}
\runauthor{Eun-Joo Kim}
\begin{document}
\maketitle
\begin{abstract}
The rapidity and centrality dependence of baryon to meson ratios
in Au$+$Au, Cu$+$Cu and p$+$p collisions at $\sqrt{s_{NN}}$ = 200 GeV
at RHIC is presented.
The $\bar{p}/\pi^{-}$ ratios are founded to be independent 
of collision system at a fixed $<N_{part}>$ at mid- and 
forward rapidities.
\end{abstract}
%
%%%%%%
%%%%%%
%
\section{Introduction}
One of the unexpected observations at RHIC was
the enhancement of proton and anti-proton yields relative to pion yields
at intermediate $p_{T}$ around midrapidity~\cite{phenix},
which was not seen in elementary hadronic collisions.
An explanation for this behavior was proposed by
the combination of pQCD calculations with soft physics 
and jet quenching~\cite{vitev1}. 
It has been  demonstrated that the fragmentation functions 
at large momentum fraction play an important role 
in hadron production~\cite{zhang}.
Alternative models that attempt to describe these baryon to meson ratios
include a phase-space determined parton coalescence picture~\cite{coal1,coal2},
as well as the dynamics driven models of recombination~\cite{hwa},
baryon-junction transport~\cite{vitev2,pop}, 
and hydrodynamic flow~\cite{hirano}. 
Central Au+Au collisions are expected to provide a partonic medium
where the coalescence of soft partons can occur.
Coalescence will depend on the system size of the partonic fluid
and is expected to be less influential for the Cu+Cu system.
In this process, the $p/\pi$ ratios can provide important information
on the dynamics of how the medium evolves longitudinally.
The BRAHMS experiment\cite{brahms_nim,brahms_white} has studied
rapidity dependent baryon to meson ratios at $y=0$ and $\eta \sim 3.2$  
for Au$+$Au, Cu$+$Cu, and p$+$p collisions at  $\sqrt{s_{NN}}$ = 200 GeV.
\section{Results}
\begin{figure}[htb]
\begin{minipage}[t]{70mm}
\centering
\includegraphics[width=7.0cm,keepaspectratio]{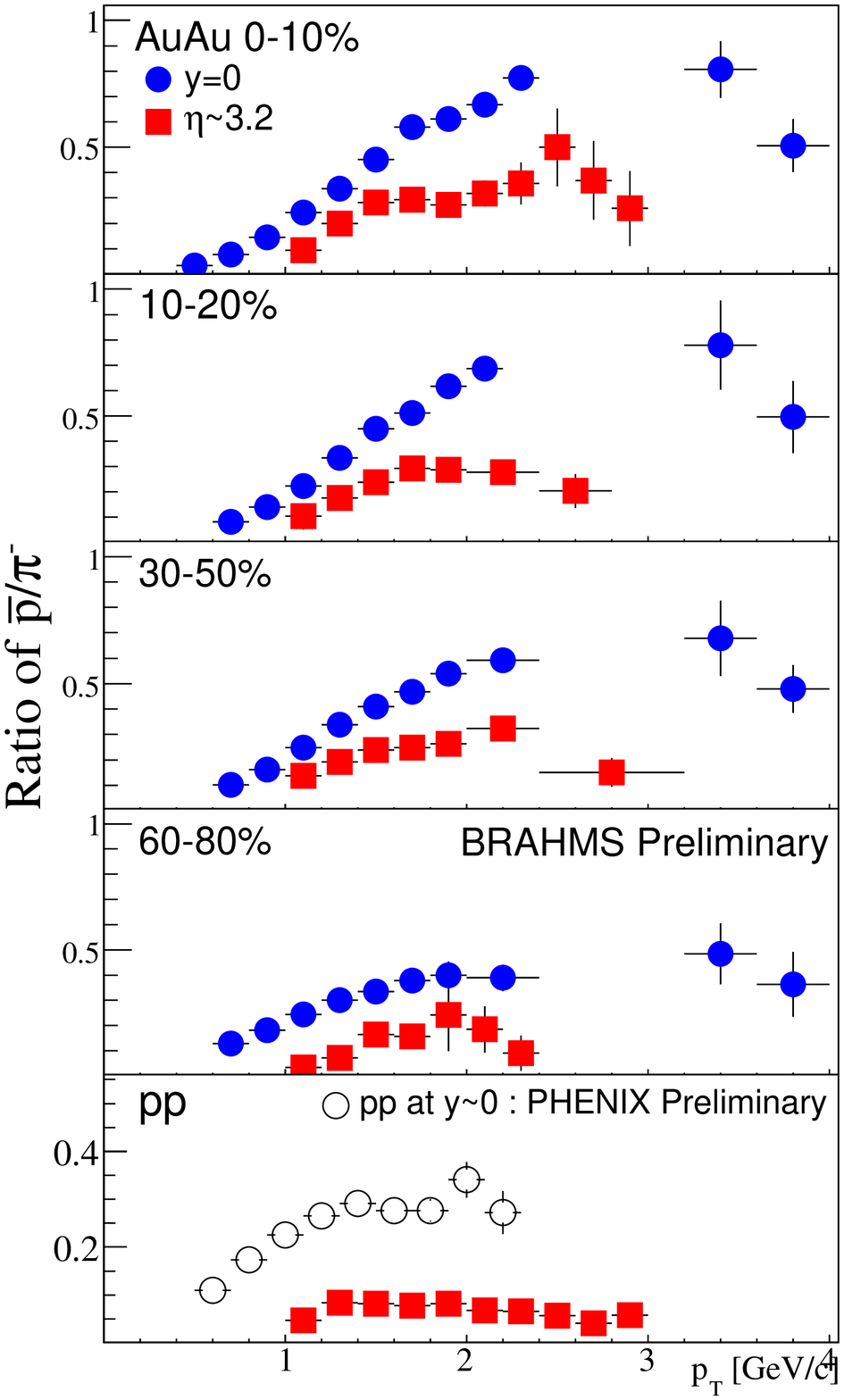}
\caption{Centrality dependent $\bar{p}/\pi^{-}$ ratios 
	 at $y=0$ and $\eta \sim 3.2$
         in Au$+$Au collisions at $\sqrt{s_{NN}}$ = 200 GeV.
         Data at p$+$p collisions at $\sqrt{s_{NN}}$ = 200 GeV are also shown.
         No $\bar{\Lambda}$ feed-down correction applied.
         Error bars are statistical only.}
\label{cent_dep}
\end{minipage}
\hspace{\fill}
\begin{minipage}[t]{85mm}
\centering
\includegraphics[width=8.5cm]{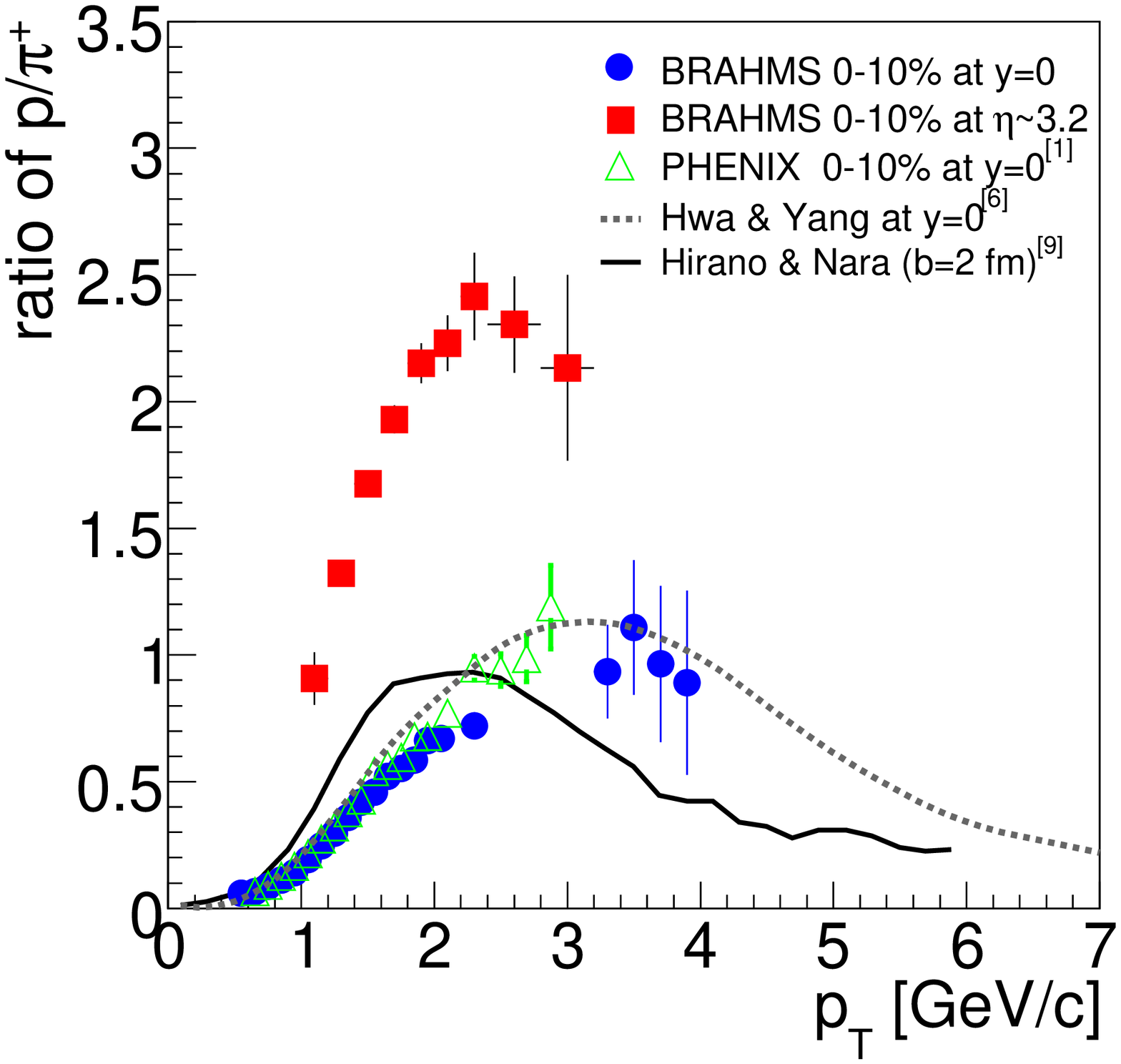}
\caption{Preliminary $p/\pi$ for 0-10\% central Au$+$Au collisions 
         at rapidity $y=0$ and $\eta \sim 3.2$. 
	 Feed-down corrections are applied at $y=0$,
         but not at $\eta \sim 3.2$.
         Comparisons with model calculations at $y=0$ are shown by curves.}
\label{model_comparison}
\end{minipage}
\end{figure}

The data were obtained with two movable BRAHMS spectrometer arms, 
the Mid-Rapidity Spectrometer~(MRS) at $90^{\circ}$, 
and the Forward Spectrometer~(FS) at $4^{\circ}$,
and global detectors for event characterization.
Particle identification was done using
time-of-flight and threshold gas Cherenkov measurements in the MRS, 
and a ring imaging Cherenkov detector~(RICH) for the FS.
Figure~\ref{cent_dep} shows the $\bar{p}/\pi^{-}$ ratios
as a function of $p_{T}$ obtained at midrapidity and forward rapidity
for the 0-10\%, 10-20\%, 30-50\% and 60-80\% centrality bins 
in Au$+$Au collisions at $\sqrt{s_{NN}}$ = 200 GeV.
The data for the same energy p$+$p collisions are also shown.
The antiproton to $\pi^{-}$ ratios show a smooth increase 
from peripheral to central collisions, 
and the data for peripheral Au$+$Au collisions approach the p$+$p results.
The centrality dependence is stronger at midrapidity than at forward rapidity,
and the peak in the $\bar{p}/\pi^{-}$ ratio is lower at forward rapidity
as compared to midrapidity.
%
%
%%%%%%%%%%%%%%%
%% Figure 1,2
%%%%%%%%%%%%%%%
Theoretical model calculations for $p/\pi$ ratios are compared to 
the experimental data in Fig.~\ref{model_comparison}.
Parton coalescence~\cite{coal1} and recombination~\cite{hwa} models
describe the observed ratios well at midrapidity, 
but three-dimensional~(3-D) hydrodynamic model~\cite{hirano}
also reproduce the observed features of $p/\pi$ enhancement 
without depending on baryon junction or coalescence.
In the 3-D hydrodynamic model, the interplay between soft and hard reaction 
components is expected to occur at intermediate $p_{T}$, and the model defines
a crossing point in transverse momentum, $p_{T,cross}$, which
might be related to the peaks evident in Fig.~\ref{cent_dep}. 
The $p/\pi$ ratios are enhanced at forward rapidity, 
as shown in Fig.~\ref{model_comparison}.
%The current hydro calculations do not show a rapidity dependence
%since baryon or isospin chemical potentials are not
%included in their schemes.  
Even though protons at high rapidity are expected to develop 
mostly from the projectile,
rapidity dependent recombination and/or radial flow effects
at the partonic level, or the inclusion of a varying baryo-chemical potential
in the hydro calculations, will likely be needed to fully understand 
the experimental observations.
%
%%%%%%%%%%%%%
%% Figure 3
%%%%%%%%%%%%%
\begin{figure}[htb]
\centering
\includegraphics[width=9.1cm,keepaspectratio]{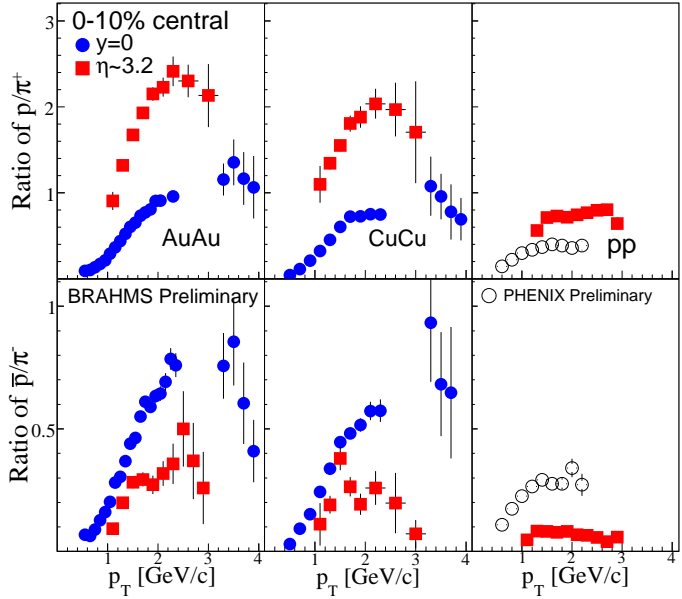}
\caption{$p/\pi$ ratios as a function of $p_{T}$ for different systems, 
         Au$+$Au, Cu$+$Cu, and p$+$p collisions 
         at midrapidity and forward rapidity.
 	 No $\Lambda$ and $\bar{\Lambda}$ feed-down corrections applied,
	 and the error bars are statistical only.}
\label{sys_dep}
\end{figure}

The $p/\pi$ ratios for different collision systems as a function of 
transverse momentum at $y=0$ and $\eta\sim3.2$
are shown in Fig.~\ref{sys_dep}.
The proton to pion ratios in $p_{T}$ increase with
the colliding system size from p$+$p to Au$+$Au collisions.
The peaks in the data, presumably related to crossing points
in $p_{T}$, are at similar positions 
for Au$+$Au and Cu$+$Cu collisions
at the same rapidity. 
%%%%%%%%%%%%%
%% Figure 4
%%%%%%%%%%%%%
\begin{figure}[htb]
\centering
\includegraphics[width=9.5cm,keepaspectratio]{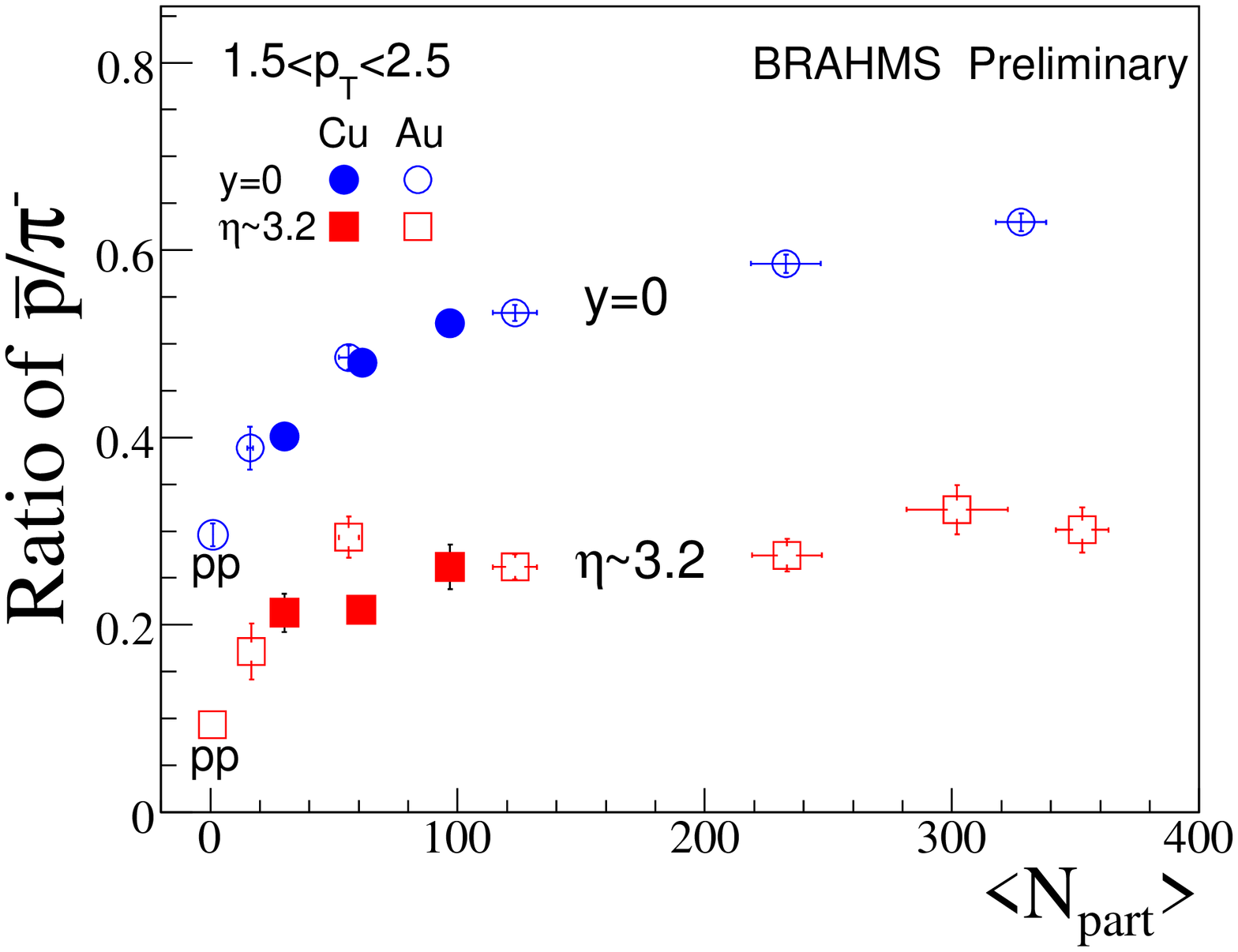}
\caption{Rapidity and $<N_{part}>$ dependence 
of $\bar{p}/\pi^{-}$ ratios for $1.5<p_{T}<2.5$ 
at $y=0$ and $\eta \sim 3.2$ 
in Au$+$Au and Cu$+$Cu collisions at $\sqrt{s_{NN}}$ = 200 GeV.
Circle(closed) symbols are the results at $y=0$,
and square(closed) symbols are the results at $\eta \sim 3.2$ 
in Au$+$Au (Cu$+$Cu) collisions. 
The error bars are statistical only.
}
\label{dep_npart}
\end{figure}

Figure~\ref{dep_npart} shows the centrality dependence
of $\bar{p}/\pi^{-}$ ratios as a function of the number of
participating nucleons, $<N_{part}>$,
at different rapidities, $y=0$ and $\eta \sim 3.2$, for
Au$+$Au and Cu$+$Cu systems at $\sqrt{s_{NN}}$ = 200 GeV.
The integrated $\bar{p}/\pi^{-}$ ratios with $1.5<p_{T}<2.5$
are shown for each centrality bin.
The data indicate that the particle ratios increase for both rapidities
going to more central collisions
up to $N_{part}$ $\approx$ 100, and show a weak dependence
from the midcentral to the most central collisions.
The $\bar{p}/\pi^{-}$ ratios with $N_{part}$ 
from different colliding systems 
are independent of system size at both rapidities.
\section{Summary}
In summary,
BRAHMS has measured rapidity dependent baryon to meson ratios 
for Au$+$Au and Cu$+$Cu at $\sqrt{s_{NN}}$ = 200 GeV
as a function of $p_{T}$ and centrality. 
The ratios are enhanced in nucleus-nucleus collisions
compared to p$+$p collisions and increase with centrality.
Positive and negative ratios show similar interplay 
between soft and hard processes for different systems.
The $\bar{p}/\pi^{-}$ ratios for different colliding systems show
a similar $<N_{part}>$ dependence at a fixed rapidity,
and no significant changes for the baryon-meson production mechanism
are observed with different beam species.
\section{Acknowledgments}
This work was supported by the division of Nuclear Physics of the
Office of Science of the U.S. Department of Energy, 
the Danish Natural Science Research Council, 
the Research Council of Norway, 
the Polish State Committee for Scientific Research 
and the Romanian Ministry of Education and Research.

\end{document}